\documentclass[prl,twocolumn,floats,aps,epsfig,nofootinbib,axodraw,amssymb]{revtex4}

\def\beq{\begin{equation}}
\def\eeq{\end{equation}}
\def\bea{\begin{eqnarray}}
\def\eea{\end{eqnarray}}

\usepackage{graphicx}
\usepackage{bm}
\begin{document}

\bibliographystyle{apsrev}

\renewcommand{\thefootnote}{\arabic{footnote}}

\title{Stability in MaVaN Models}

\author{Christopher Spitzer\footnote{Email: cspitzer@u.washington.edu}}
\affiliation{Department of Physics, Box 1560, University of Washington}

\begin{abstract}
Mass-varying neutrino (MaVaN) models propose a source of dark energy in a new
scalar field called the acceleron.  Recent work has shown that nonrelativistic
neutrino fields in these theories are unstable to inhomogeneous fluctuations,
and form structures that no longer behave as dark energy.  One might expect that
in multiple-neutrino models, the lighter species could continue to act as a source
for the acceleron, generating dark energy without the help of heavier species.
This paper shows that by considering the evolution of the acceleron field for a
large class of models, the result of any component becoming unstable is that all
components become unstable within a short time on cosmological scales.  An alternate
model employing a second scalar field in a hybrid potential is shown to have stable
MaVaN dark energy even in the presence of unstable heavier components.
\end{abstract}

\maketitle

\section{Introduction}

Recent results in precision cosmology have created a number of puzzles.  Among the
unexplained phenomena are several apparent coincidences in which the energy density
of two components are comparable despite having different redshift properties.  Another
is the existence of a negative pressure cosmological fluid that accelerates the rate
of expansion of the universe.  The model of mass-varying neutrinos (MaVaNs),
introduced by Fardon et al. in~\cite{Fardon:2003eh}, suggests that neutrino and
dark energy densities have tracked each other throughout the lifetime of the
universe through a new scalar field called the acceleron.  The energy density of
the scalar potential of this new field contributes to the dark energy.  The authors
showed that this new field is capable of explaining the present cosmological expansion.

Since the original MaVaN proposal, several authors have investigated the stability
of the \emph{dark sector}, composed of neutrinos and the scalar field, under perturbations
to the neutrino density~\cite{Afshordi:2005ym,Takahashi:2006jt}.  Both groups subject the
model to a hydrodynamic analysis.  In this approximate picture, the speed of sound squared
in the cosmological fluid, given by~\cite{Mukhanov:1990me},
\beq
c_s^2=\frac{\dot{P}}{\dot{\rho}}=w+\frac{\dot{w}\rho}{\dot{\rho}}=w-\frac{\dot{w}}{3H(1+w)}
\eeq
is positive only if
\beq
\frac{\partial w}{\partial z} \geq -\frac{3w(1+w)}{1+z}
\eeq
where $z$ is the redshift.  However, for nonrelativistic neutrinos,
$\frac{\partial w}{\partial z}$ is
a negative quantity while the right hand side is positive for $w$ close to -1.  Since at
least one neutrino must be nonrelativistic today, the dark sector appears unstable.

Additionally, Afshordi, et al. perform a stability analysis using kinetic theory to
account for neutrino streaming~\cite{Afshordi:2005ym}.  This analysis also
shows that perturbations in the neutrino field become unstable when the mass of the
neutrino is of order the neutrino temperature.  The ratio of mass to temperature when
the instability occurs is a function of the acceleron potential, but is approximately
7 for the potentials considered in this paper.  Afshordi, et al., also examine the result
of undergoing a phase transition in a neutrino component, and suggest that the
unstable neutrino field may rapidly form nonlinear structures termed
\emph{neutrino nuggets}, which redshift as dark matter and cannot drive the cosmic
expansion.

In this paper, we will assume that the neutrinos are initially relativistic.  As
the universe expands and cools, the neutrinos become less relativistic until
their mass and temperature are approximately equal.
At this point, we assume
that they decouple from the scalar field into some structure such as the one described
by Afshordi et al.  The resulting dark matter will not provide a large contribution
to the energy density, and we will not include the contribution in the dark sector
energy.  Note that in this paper dark sector will always refer to the energy density
contributions of the stable neutrinos and the scalar field.

In a model with multiple neutrinos, the dark energy may still be driven by relativistic
species even after the heavier components have become unstable.
However, the coupling between the neutrinos and acceleron create a feedback mechanism.
The shift in the acceleron expectation value when a neutrino becomes unstable can
be sufficient to necessarily change the mass of another species so that it goes from
relativistic to nonrelativistic, causing it to become unstable as well.  This
\emph{cascaded instability} can make all neutrinos unstable at about the same time
that the heaviest neutrino becomes unstable.
This paper will examine a particular class of models, and will show that see-saw MaVaN
models with flat scalar potentials suffer from precisely this problem.
The timing between the instability in successive neutrinos is strongly dependent on the
flatness of the scalar potential, but a flat potential is also required to generate
the observed dark energy.  This result may point toward models that do no suffer
from a cascaded instability, and this paper concludes with a simple example.

\section{See-Saw Models}

Consider a model with $n$ active neutrinos in which each neutrino is paired with
a sterile counterpart, which is coupled to a new scalar field,
\beq
-L \supset \sum_{i=1}^n (M^\prime_i \nu_i N_i + \lambda A N_i N_i)
\eeq
where $\nu_i$ are the active neutrinos, $N_i$ are the sterile neutrinos, and $A$ is
the acceleron scalar field.  $M_i$ and $\lambda$ describe coupling strengths.  If we
assume that the normal see-saw limit holds, $\langle \lambda A\rangle \gg M_i$, the
system reduces to an effective Lagrangian describing active neutrinos with a
Majorana mass term,
\beq
-L_{eff} \supset \sum_{i=1}^n \frac{M_i^2}{A} \nu_i \nu_i \equiv \sum_{i=1}^n m_i \nu_i \nu_i
\eeq
where $M_i^2=M^{\prime 2}_i/\lambda$, and $m_i$ is the effective mass.

The energy density of the dark sector is
\beq
\rho_d=\rho_\nu+V(A)
\label{eq_rho}
\eeq
where $V$ is the potential of the acceleron field.  Assuming that the neutrino
distribution function is a stretched thermal distribution, the value of the equation of
state for this dark sector is (one derivation is given in~\cite{Peccei:2004sz})
\beq
w=\frac{p_d}{\rho_d}=
\frac{T^4 \sum_i \left[ 4
   F\left(\frac{m_{i}^2}{T}\right)-
   J\left(\frac{m_{i}^2}{T}\right)\right]}{3 \left[T^4 \sum_i 
   F\left(\frac{m_{i}^2}{T}\right)
   +V(A)\right]}-1
\label{eq_w}
\eeq
where $i$ runs over the active neutrino species, and we have defined the distribution
function and its derivative,
\bea
F(x)&=&\frac{1}{\pi^2}\int_0^\infty \frac{y^2\sqrt{y^2+x^2}dy}{e^y+1}
\\
J(x)&=&\frac{x^2}{\pi^2}\int_0^\infty \frac{y^2dy}{\sqrt{y^2+x^2}(e^y+1)}
\eea
Fardon, et al., show
in~\cite{Fardon:2003eh} that the system remains very close to the minimum of the
effective potential and evolves adiabatically.  For the model above, this minimization
conditions becomes
\beq
\frac{\partial V}{\partial A}=\sum_{i=1}^n\int_0^{\infty}\frac{dy}{\pi^2}\frac{m_{i} T^2 y^2}{(y^2+\frac{m_{i}^2}{T^2})^\frac{1}{2}}\frac{1}{1+e^y}(-\frac{\partial m_{i}}{\partial A})
\label{eq_min}
\eeq
where $y=\frac{p_\nu}{T}$.

Commonly employed forms for the potential include small power law ($V=BA^k,k\ll 1$),
logarithmic ($V=Blog(A/A_0)$) and quadratic ($V=BA^2$).  After starting with the
small power law case, it will be easy to generalize to all cases by assuming
$\partial V/\partial A=BkA^{k-1}$ with $B$ and $k$ unrestricted.

\section{Approximations}

The expectation value of the acceleron at a particular value of $z$ is determined by
the minimization equation~(\ref{eq_min}).  Unfortunately, this equation in general
does not have a closed form solution.  To examine the behavior of this equation it is
useful to approximate the result in three ranges: relativistic ($m_\nu \ll T$),
quasirelativistic ($m_\nu \sim T$), and nonrelativistic ($m_\nu \gg T$).  In these
approximation, the minimization equation becomes
\bea
\label{eq_approxr}
R: & \frac{\partial V}{\partial A} \simeq \frac{1}{A^3}\frac{M_i^4 T^2}{\pi^2}I_1\\
\label{eq_approxnr}
NR: & \frac{\partial V}{\partial A} \simeq \frac{1}{A^2}\frac{M_i^2 T^3}{\pi^2}I_2\\
\label{eq_approxqr}
QR: & \frac{\partial V}{\partial A} \simeq \frac{1}{A^3}\frac{M_i^4 T^2}{\pi^2}I_3
\eea
with the unitless $O(1)$ integrals
\bea
I_1&=&\int_0^\infty dy \frac{y}{1+e^y} \simeq 0.822 \\
I_2&=&\int_0^\infty dy \frac{y^2}{1+e^y} \simeq 1.803 \\
I_3&=&\int_0^\infty dy \frac{y^2}{(1+e^y)\sqrt{1+y^2}} \simeq 0.670
\eea

\begin{figure}[tbh]
\includegraphics[width=8cm]{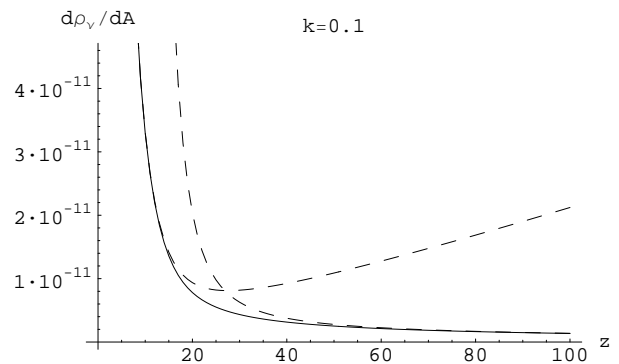}
\caption{The derivative of neutrino energy density with respect to A.  The solid line is a numerically determined result, and the dashed lines are approximations in the nonrelativistic (low z) and relativistic (high z) regimes.}
\label{fig_approx1}
\end{figure}

A comparison of these approximations to unapproximated numerical simulation for the two
neutrino models discussed in the next section is shown in in figure~\ref{fig_approx1}.
This figure shows the contribution of the derivative of the neutrino term to
the minimization equation, which are the right hand sides of
equations~(\ref{eq_approxr}-~\ref{eq_approxnr}).
The relativistic and nonrelativistic approximations (dashed curves) are a good
match to the numerically calculated value in their respective regions of validity
at high and low redshift.

\section{Two Active Neutrinos, Small Power Law}

As a warm-up, consider the case of two active neutrinos, with $m_1 \gg m_2$, and
a power law potential with $0<k\ll1$.  To first order, the derivative of the potential
becomes (accurate to 10\% for typical small values of $k$)
\beq
\frac{\partial V}{\partial A} \simeq \frac{B k}{A}
\eeq

At large $z$, both neutrinos will be relativistic.  As the universe cools, the neutrino
temperature decreases while mass increases, and at some value of $z$ neutrino 1 becomes
quasirelativistic, and will become unstable.  Due to the original mass hierarchy,
neutrino 2 is still relativistic at this point.

Now consider the acceleron expectation value and neutrino masses both before and
after the transition of neutrino 1 to a dark matter phase.  Before the transition,
when neutrino 1 has mass of order the temperature, we have
\bea
m_{1,before} \sim T  \rightarrow A_{before} \sim \frac{M_1^2}{T} \\
m_{2,before}=\frac{M_2^2}{A_{before}} \sim \frac{M_2^2}{M_1^2}T 
\eea
So our mass hierarchy implies $M_2^2 \ll M_1^2$.

The minimization equation~(\ref{eq_min}), with the approximations from~(\ref{eq_approxr})
and~(\ref{eq_approxqr}), yields the value of the acceleron prior to the
transition,
\beq
A_{before}=\frac{T}{\pi}\frac{1}{\sqrt{Bk}}\sqrt{M_1^4 I_3 + M_2^4 I_1}
\eeq

Since the acceleron will suddenly change value when neutrino 1 becomes unstable and
stops sourcing it, we do not know a priori whether neutrino 2 will be R, NR or QR after
the transition.  By trying each in turn, we quickly find that only the QR
assumption is consistent.  For instance, an assumption that neutrino 2 stays
relativistic yields a value for $m_2$ that is of order the temperature, which
violates the assumption.  Looking carefully at the QR case, from the $m_1 \sim T$
condition before the transition, we have the relationship
\beq
\frac{\sqrt{B k}}{T}\frac{\pi}{\sqrt{I_3}} \sim T
\eeq
which provides the temperature at the time of transition in terms of the parameters
of the system.
Solving for the acceleron and neutrino 2 mass after the transition, we have
\bea
A_{after}=\frac{M_2^2 T}{\pi}\sqrt{\frac{I_1}{Bk}} \\
m_{2,after}=\frac{\sqrt{Bk}}{T}\frac{\pi}{\sqrt{I_3}} \sim T
\eea

The picture that emerges from this small example is that after the first neutrino becomes
unstable and decouples from the acceleron, the acceleron assumes a value
that pushes the second neutrino into a quasirelativistic region.  Once this occurs,
the second neutrino will also soon become unstable.  In the end there is nothing left
to drive the dark energy.

\section{Generalized Models}

The simple two-neutrino, small-power law model generalizes easily to include a larger
number of neutrino species and different potentials.  Additional neutrinos
do not improve the stability picture, but decreasing the flatness of the potential does.

First consider the addition of a third neutrino that is much less massive than
neutrino 1.  This requires $M_2^2 \ll M_1^2$ and $M_3^2 \ll M_1^2$.
After the transition, the value of the acceleron becomes
\beq
A_{after}=\frac{1}{\sqrt{Bk}}\frac{T}{\pi}\sqrt{M_2^4 I_1 + M_3^4 I_1}
\eeq
As a result, the mass of the second and third neutrinos become
\bea
m_{2,after}=\sqrt{\frac{I_3}{I_1}}\frac{M_2^2}{\sqrt{M_2^4 + M_3^4}}T \\
m_{3,after}=\sqrt{\frac{I_3}{I_1}}\frac{M_3^2}{\sqrt{M_2^4 + M_3^4}}T
\eea
Neutrino 2 is relativistic only if $M_2 \ll M_3$, but similarly neutrino 3
is relativistic only if $M_3 \ll M_2$.  Since we can not satisfy both these conditions,
at least one of the remaining neutrinos is quasirelativistic, and becomes unstable.

Using similar arguments, it is easy to see that for the general case of $n$ neutrinos,
all neutrinos will become unstable within a short period of each other.

Now consider a more general potential, $V=BA^k$ (for an arbitrary k), for the
two-neutrino case.  The mass of neutrino 2 after neutrino 1 becomes unstable is
\beq
m_{2,after} \simeq \left(\frac{I_3}{I_1}\right)^{\frac{1}{k+2}} M_1^{\frac{-2k}{k+2}} M_2^{\frac{2k}{k+1}} T
\eeq
In this case, to keep neutrino 2 relativistic, we require
\beq
\left(\frac{M_2}{M_1}\right)^{\frac{2k}{k+2}} \ll 1
\label{eq_kcond}
\eeq
Note that this does allow the light neutrino to continue to drive the acceleron,
but only if the acceleron potential is flatter than the small-$k$ power we examined above.
Unfortunately, this potential predicts a dark energy equation of state that is in
conflict with observation.  The value of $k$ is connected to the present value of the
equation of state by
\beq
k=-\frac{1-w_0}{w_0}
\eeq
Requiring a value of $w_0$ close to $-0.9$ yields a value of $k$ that lies in the region
where the lighter neutrinos become unstable very quickly.  Conversely, values of $k$ that
are large enough to keep the lighter neutrinos relativistic also predict a value of $w_0$
too large.

\section{Numerical Simulation}

\begin{figure}[tbh]
\includegraphics[width=8cm]{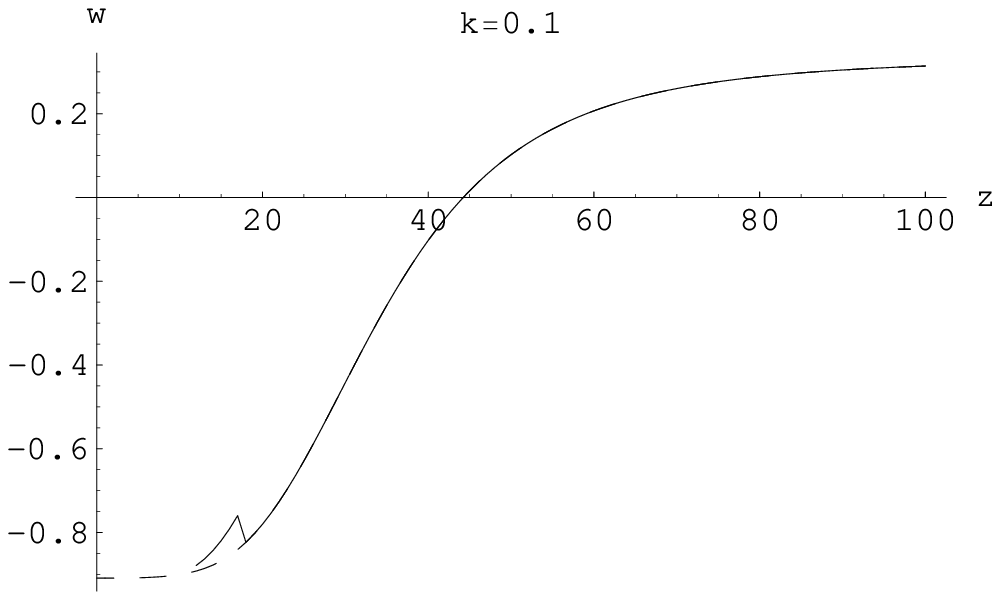}
\includegraphics[width=8cm]{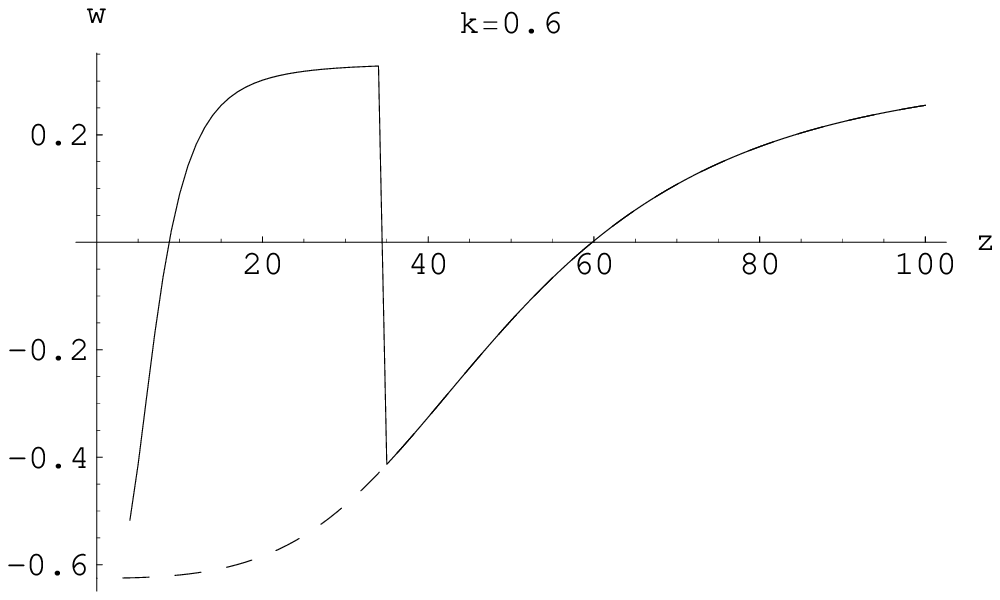}
\caption{Simulation of the dark sector (neutrinos and dark energy) equation of state of a two neutrino system under the power law with two values of k.  The dashed line is the result if instability is ignored, while the solid line removes an unstable component from the dark sector.  The solid lines end when all neutrinos have become unstable.  The contributions of the dark matter formed by unstable neutrino components is not included here.}
\label{fig_ev1}
\end{figure}

We verified the above relationships using an unapproximated numerical simulation.
An example of the dependence of the stability of lighter neutrinos on the steepness
of the acceleron potential is shown in figure~\ref{fig_ev1}.  The plots shows the
evolution of the equation of state for the neutrino and dark energy components, as
given in~(\ref{eq_w}), calculated numerically, of a two neutrino system for
two values of $k$.  Note that this $w$ does not include contributions from either
the dark matter terms that resulted from unstable neutrino components, or from components
that were not included in the dark sector as defined above.  As $k$ increases, the
lighter neutrino is longer-lived, and can continue to drive the dark energy closer to
today ($z=0$).  However, this also results in $w$ approaching a disallowed value.

Comparing the set of approximations made in simplifying the minimization equation
to the numerical results show that they are accurate to 15-50\%.  Noting that the
order-of-magnitude arguments above already contain multiplicative $O(1)$ corrections,
the contribution of numerical inaccuracy does not affect our conclusions.

\section{Hybrid Model}

The condition for the stability of lighter neutrino species, given in
equation~(\ref{eq_kcond}), suggests that we need to find a model that does not require
a flat potential to generate dark energy.  There are several ways to
achieve this, and here we will discuss a particularly simple extension.  This new
model employs a second scalar field that is not directly coupled to the neutrinos
but provides large contribution to the dark energy.  Since the configuration of
the potential is borrowed from hybrid inflation (see~\cite{Linde:1993cn}), the new model
is called the ``Hybrid MaVaN'' model.

The hybrid model includes a scalar field, $\sigma$, that acts as the
\emph{waterfall field}.  The potential is
\beq
V = b^2A^2+g^2A^2\sigma^2+(h^2-\alpha\sigma^2)^2
\eeq
where $b$,$g$,$h$ and $\alpha$ are new coupling constants.  The precise details of
the potential are unimportant, as long as it supports a false minimum as discussed
below.

The hybrid potential has two distinct regions of behavior under variation
of $\sigma$.  If $2\alpha h^2>g^2A^2$, then there are two minima at
$\pm\sqrt{\frac{2\alpha h^2-g^2A^2}{2\alpha^2}}$.  Otherwise, there is a single
``false minimum'' at $0$.  Forcing the field into this false minimum by requiring
\beq
A>\sqrt{2\alpha}h/g
\eeq
the potential becomes
\beq
V \rightarrow b^2A^2+h^4
\label{eq_hybridv}
\eeq
which includes a cosmological-constant type term $h^4$.  This term dominates the
acceleron contribution to the potential if $h\gg \sqrt{2\alpha}b/g$,
and from  equations~(\ref{eq_rho}) and~(\ref{eq_w}) may also dominate over the neutrino
contribution to the energy density.  If we also assume the
see-saw condition, $A\gg m_i/\lambda$, then the model described in previous sections
can be used without any change other than using the form of the potential in
equation~(\ref{eq_hybridv}).

\begin{figure}[tbh]
\includegraphics[width=8cm]{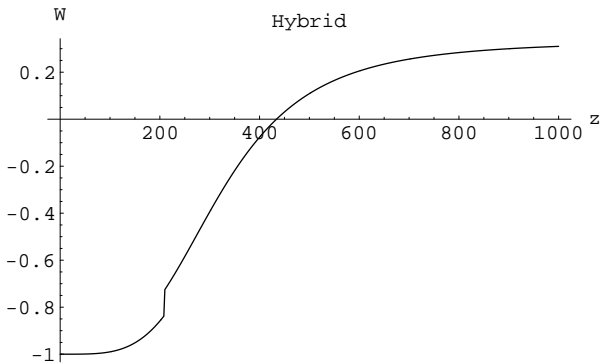}
\caption{Simulation of the equation of state of a three neutrino hybrid potential system.  The kink at $z=200$ occurs when the heaviest neutrino becomes unstable.  The remaining two neutrinos continue to be relativistic and stable until z=0.}
\label{fig_hybridw1}
\end{figure}

The quadratic dependence on $A$ in equation~(\ref{eq_hybridv}) means that the stability
condition for the lighter neutrinos in equation~(\ref{eq_kcond}) is easily satisfied.
The allowed parameter range is quite large, and it is straightforward to find coefficients
that are produce observationally allowed values of neutrino mass and equation of state.
The numerical simulation of the evolution of one such model, with a hierarchy of masses
and $h=0.06$ is shown in figure~\ref{fig_hybridw1}.  The lighter two neutrinos stay
relativistic until $z=0$, despite the instability in the massive neutrino.
Note this model achieves both the stability of the lighter neutrinos and has a dark
sector equation of state of $w=-1$ at $z$ near $0$.

\section{Conclusion and Discussion}


Several authors have found that a nonrelativistic MaVaN neutrino field is unstable
to inhomogeneous fluctuations.  By considering the evolution of the acceleron expectation
value
when a neutrino field becomes unstable, we have argued above that all neutrino fields
in the theory are susceptible to a cascaded instability in which they all become
unstable at nearly the same time.  This occurs as long as the scalar potential has
a nearly flat dependence on the acceleron, and the neutrino masses vary inversely with
the acceleron.  Including a very light neutrino is not sufficient to avoid this problem.
Since there are at least three neutrino species, and the atmospheric neutrino
deficit requires at least one mass scale above 1eV, the instability poses a constraint
on all physical MaVaN models.

There are a number of possible resolutions.  One is to increase the curvature of
the scalar potential.  In models with a single scalar field, this makes it difficult
for the scalar field potential to form dark energy.  However, this is easily remedied
by including a second scalar field.  A simple example is illustrated above in the
hybrid MaVaN model.  A similar potential arises naturally in supersymmetric models,
such as those in~\cite{Fardon:2005wc} and~\cite{Takahashi:2005kw}.

Another solution is to modify the theory so that the dark sector never reaches a state
where the adiabatic condition applies.  Such models do not suffer from the instability
described above.  One such theory is presented in~\cite{Brookfield:2005td}.

Reducing the dependence of the acceleron on the heavy neutrino components also
forms a class of possible solutions.  If the acceleron is decoupled from each heavy
component before it becomes unstable, the acceleron expectation value is not
quickly driven to a new scale.  The mass of the lighter neutrinos would be largely
unaffected, avoiding instability.

\section{Acknowledgments}

I would like to thank Ann Nelson for useful conversations and guidance while
working on this project.  This work was supported in part by the Department of
Energy.

\bibliography{mavan}

\end{document}